 \documentstyle[aas2pp4,epsf]{article}

\newcommand{\Ell}{E_\parallel}      
\newcommand{\sgP}{\sigma_{\rm p}}  
\newcommand{\rlc}{\varpi_{\rm LC}} 
\newcommand{\Ex}{\epsilon_{\rm x}} 
\newcommand{\Eg}{\epsilon_\gamma}  
\newcommand{\inc}{\alpha_{\rm i}}  

\begin{document}

\title{ELECTRODYNAMIC STRUCTURE OF OUTER GAP ACCELERATORS:
       INVISIBILITY OF PULSED TEV FLUXES}
\author{K. Hirotani}
\affil{National Astronomical Observatory, 
       Mitaka, Tokyo 181-8588, Japan;
       hirotani@hotaka.mtk.nao.ac.jp}
\authoremail{hirotani@hotaka.mtk.nao.ac.jp}

\begin{abstract}
We study the structure of an outer-magnetospheric gap
around a rotating neutron star.
Migratory electrons and positrons are accelerated by the
electric field exerted in the gap 
and radiate copious gamma-rays via curvature process.
Some of the gamma-rays materialize 
by colliding with the X-rays illuminating the gap, 
leading to a pair production cascade. 
The replenished charges partially screen the original acceleration field, 
which is self-consistently solved from the Poisson equation,
together with the Boltzmann equations for gamma-rays and
the continuity equations for particles.
We demonstrate that it is difficult to detect the TeV emission
due to Compton upscatterings in the gap,
by the current ground-based telescopes.
\end{abstract}

\keywords{gamma-rays: observation -- gamma-rays: theory -- 
          magnetic field -- X-rays: observation}


\section{Introduction}

In the very high energy (VHE) region above 100 GeV,
positive detections of radiation at a high confidence 
level have been reported from the direction of the 
Crab, B1706-44, and Vela pulsars 
(Bowden et al. 1993; Nel et al. 1993; Edwards et al. 1994;
 Yoshikoshi et al. 1997; 
 see also Kifune 1996 for a review),
by virtue of the technique of imaging Cerenkov light
from extensive air showers.
However, as for {\it pulsed} TeV radiation,
only the upper limits have been obtained from these pulsars.
If the VHE emission originates the pulsar magnetosphere,
rather than the extended nebula,
significant fractions of them are expected to show a pulsation.
Therefore, the lack of {\it pulsed} TeV emissions provides a
severe constraint on the modeling of particle acceleration zones
in pulsar magnetospheres.

In the picture of
Cheng, Ho, and Ruderman (1986a,b),
who developed a version of an outer magnetosphere $\gamma$-ray emission 
zone,
the magnetosphere should be optically thick for pair production
in order that the TeV flux is absorbed to be unobservable.
This, in turn, requires very high luminosities of infrared photons,
which are generally orders of magnitude
larger than the observed values (Usov 1994).
We are therefore motivated by the need to contrive an outer gap model
which produces less TeV emission with a moderate infrared luminosity.

More recently, Romani (1996) showed that the pulsed TeV flux
can be somewhat less than $1\%$ of the pulsed GeV flux,
assuming the functional form of the acceleration field 
as $\Ell \propto 1/r$, where $r$ denotes the distance from the star.
Subsequently, Hirotani (1999, Paper IV) revealed semi-analytically
that the TeV fluxes from outer gaps are unobservable,
by considering a \lq gap closure condition' so that 
a single pair produces copious $\gamma$-rays one of which materialize
as a pair on average
(see also Zhang \& Cheng 1998 for another kind of closure condition).
In this letter, instead of imposing the closure condition,
we investigate the spatial distribution of $\Ell$,
by solving the Poisson equation and the Boltzmann equations 
simultaneously.
This method was first developed by Beskin et al. (1992)
for black-hole magnetospheres and applied to
pulsar magnetospheres by  
Hirotani \& Shibata (1999a,b,c; hereafter Paper I, II, III).

We formulate basic equations in \S~2, demonstrate 
the invisibility of pulsed TeV fluxes in \S~3,
and discuss the validity of assumptions in \S~4.

\section{Basic Equations and Boundary Conditions}

\subsection{Reduction of Poisson Equation}
\label{sec:Poisson}

To simplify the geometry, let us introduce a rectilinear coordinates
for a region around the null surface.
Suppose that the magnetic field lines are straight lines 
parallel to the $x$ axis,
where $x$ increases outwardly.
We approximate that the null surface ($x=0$) 
is perpendicular to the $x$ axis.
It is convenient to non-dimensionalize the length scales 
by a typical Debey scale length $c/\omega_{\rm p}$ 
(see the discussion in \S 3.1 in Paper~I for details), where
\begin{equation}
  \omega_{\rm p} = \sqrt{ \frac{4\pi e^2}{m_{\rm e}}
	                  \frac{\Omega B}{2\pi ce} }
	         = 1.875 \times 10^7 \Omega_2{}^{1/2} B_5{}^{1/2}
		   \mbox{rad s}^{-1};
  \label{eq:def-omegap}
\end{equation}
$e$ refers to the magnitude of the charge on the electron,
$\Omega_2$ the angular frequency ($\Omega$) of the neutron star
in $10^2 \mbox{rad s}^{-1}$ unit,
$c$ the velocity of light,
and $B_5$ the magnetic field strength ($B$) at $x=0$ in
$10^5$ G unit.
Then the dimensionless coordinate variable becomes
\begin{equation}
  \xi \equiv \frac{\omega_{\rm p}}{c} x
      = 6.25 \times 10^{-4} \Omega_2{}^{1/2} B_5{}^{1/2} x .
  \label{eq:def-xi}
\end{equation}

Assuming the trans-field thickness is comparable to or greater than
the longitudinal width,
we Fourier-analyze the Poisson equation 
in the trans-field directions to obtain
(Paper I,II,III)
\begin{equation}
  E_\parallel = -\frac{d\varphi}{d\xi},
  \label{eq:basic-1}
\end{equation}
\begin{equation}
  \frac{dE_\parallel}{d\xi}
  = -\frac{\varphi}{\Delta_\perp{}^2} + n_+(\xi) - n_-(\xi) 
    - \frac{2\pi c}{\Omega B} \rho_{\rm GJ}(\xi),
  \label{eq:basic-2}
\end{equation}
where $ \Delta_\perp \equiv (\omega_{\rm p}/c) D_\perp$ and 
$ \varphi(\xi) \equiv e\Phi(x)/(m_{\rm e}c^2)$;
$D_\perp$ is a typical gap thickness in the trans-field directions
and $\Phi$ the electrostatic potential.
The particle densities are normalized in terms of a
typical value of the Goldreich--Julian density as
\begin{equation}
  n_\pm(\xi) \equiv 
    \frac{2\pi ce}{\Omega B} N_\pm (x),
  \label{eq:def-n}
\end{equation}
where $N_\pm(x)$ are the spatial number density of $e^\pm$'s.
We evaluate it with the Newtonian value as 
\begin{equation}
  \frac{2\pi c}{\Omega B} \rho_{\rm GJ}
  = \frac{ 2\cos\theta \cos(\theta-\inc)
           -\sin\theta \sin(\theta-\inc) }
         {\sqrt{1+3\cos^2\theta} \left[ 1-(\varpi/\rlc)^2 \right]},
  \label{eq:rhoGJ}
\end{equation}
where $\alpha_{\rm i}$ refers to the inclination angle of the magnetic 
moment, $\theta$ the clatitude angle of the position at which
$\rho_{\rm GJ}$ is measured,
$\varpi$ the distance from the rotation axis,
and $\rlc \equiv c/\Omega$ the light-cylinder radius.

\subsection{Boltzmann equations}
\label{sec:contEQ}

Let us first introduce the following dimensionless $\gamma$-ray 
densities in the dimensionless energy interval
between $\beta_{i-1}$ and $\beta_i$:
\begin{equation}
  g_\pm{}^i(\xi) \equiv 
    \frac{2\pi ce}{\Omega B}
    \int_{\beta_{i-1}}^{\beta_i} d\epsilon_\gamma G_\pm(x,\epsilon_\gamma),
  \label{eq:def-g}
\end{equation}
where $G_\pm(x,\Eg)$ are the distribution functions of
$\gamma$-ray photons having momentum $\pm m_{\rm e}c \epsilon_\gamma$ 
In this letter, we set $\beta_0=10$;
therefore, the lowest $\gamma$-ray energy considered is $10 m_{\rm e}c^2$.
In this letter, 
we divide the $\gamma$-ray spectra into $13$ energy bins as follows:
$\beta_1= 30$, 
$\beta_2= 100$, 
$\beta_3= 200$, 
$\beta_4= 300$, 
$\beta_5= 600$, 
$\beta_6= 10^3$, 
$\beta_7= 2 \times 10^3$, 
$\beta_8= 3 \times 10^3$, 
$\beta_9= 6 \times 10^3$, 
$\beta_{10}= 10^4$, 
$\beta_{11}= 2 \times 10^4$, 
$\beta_{12}= 3 \times 10^4$, 
$\beta_{13}= 10^5$.

Assuming for simplicity that both electrostatic and 
curvature-radiation-reaction forces cancel out each other,
we obtain the following continuity equations 
for particles (eqs. [31] and [32] in Paper III):
\begin{equation}
  (1-q)\frac{dn_\pm}{d\xi} = 
    \pm \sum_{i=1}^{14} [ \eta_{\rm p+}{}^i g_+^i(\xi)
                         +\eta_{\rm p-}{}^i g_-^i(\xi)],
  \label{eq:basic-3}
\end{equation}
\begin{equation}
 q \equiv \frac12 \left( \frac{\pi}{2}\frac{\varpi}{\varpi_{\rm LC}}
                  \right)^2,
 \label{eq:def_q}
\end{equation}
The factor $1-q$ denotes the projection effect due to the 
three-dimensional motion of the particles,
which was not considered in Papers I, II, and III.
Since the electric field is assumed to be positive in the gap,
$e^+$'s (or $e^-$'s) migrate outwards (or inwards).

The X-ray field affects the gap structure
through the pair-production rate $\eta_{{\rm p}\pm}^i$.
Putting $\epsilon_i \equiv (\beta_i +\beta_{i-1})/2$, 
we can evaluate $\eta_{{\rm p}\pm}^i$ as
\begin{equation}
  \eta_{{\rm p}\pm}^i
  = \frac{c}{\omega_{\rm p}}
     \int_{\epsilon_{\rm th}}^\infty d\epsilon_{\rm x}
     \frac{dN_{\rm x}}{d\Ex} 
     \sgP(\epsilon_{\rm i}, \Ex,\mu_{\rm c}),
  \label{eq:def_etap}
\end{equation}
where $dN_{\rm x} / d\Ex$ denotes the number density of the X-rays
between energies $\epsilon_{\rm x}$ and $\epsilon_{\rm x}+d\epsilon_{\rm x}$;
$\sgP$ is the pair production cross section
when two photons collide with an angle $\cos^{-1}\mu_{\rm c}$
(Berestetskii et al. 1989; eqs. [21]-[23] in Paper II).
For surface blackbody X-rays,
we obtain $\mu_{\rm c}= \cos\phi_{\rm abb}\sin\theta_{\rm c}$,
where $\phi_{\rm abb}=\tan^{-1}(\varpi / \rlc)$
is the aberration angle.
For power-law, magnetospheric X-rays,
we assume $\mu_{\rm c}=\cos(0.5W/\rlc)$, 
where $W$ is the full gap width,
for both inwardly and outwardly propagating $\gamma$-rays.
As the lower bound of the integral,
the threshold energy,
$\epsilon_{\rm th} \equiv 2 (1-\mu_{\rm c})^{-1} \epsilon_\gamma{}^{-1}$, 
for pair production appears.

A combination of the two continuity equations (\ref{eq:basic-3})
gives the current density $j_0 \equiv n_+(\xi) + n_-(\xi)$,
which is constant for $\xi$.
When $j_0=1.0$, the current density equals the typical Goldreich--Julian 
value, $\Omega B / (2\pi)$.

We next consider the Boltzmann equations for $\gamma$-rays.
Integrating the $\gamma$-ray Boltzmann equations
between $\beta_{i-1}$ and $\beta_i$, we obtain (eq.~[39] in Paper~III)
\begin{equation}   
  \pm (1-q) \frac{d}{d\xi} g_\pm{}^i(\xi)
     = - \eta_{{\rm p}\pm}{}^i g_\pm{}^i(\xi)  
       + \eta_{\rm c}^i n_\pm(\xi),
  \label{eq:basic-5}
\end{equation}
where $i=1,2,\cdot\cdot\cdot,m(=13)$.
The $\gamma$-ray production rate via curvature radiation,
$\eta_{\rm c}{}^i$,
are defined by equation (40) in Paper III.
The particle motion is assumed to be mono-energetic 
and the Lorentz factor saturates in a balance between electrostatic
acceleration and the curvature drag force (see eq. [19] in Paper III).

\subsection{Boundary Conditions}
\label{sec:BD}

In the same manner as in Papers I, II, and III,
we impose the boundary conditions
at the inner ($\xi=\xi_1$) and outer ($\xi=\xi_2$) boundaries as follows:
\begin{equation}
  \Ell(\xi_1)=0, \quad  \varphi(\xi_1) = 0,
  \label{eq:BD-1}
\end{equation}
\begin{equation}
  g_+{}^i(\xi_1)=0,  \quad 
  n_+(\xi_1)=0,      \quad
  n_-(\xi_1)=j_0,
  \label{eq:BD-2}
\end{equation}
\begin{equation}
  \Ell(\xi_2)=0,    \quad
  g_-{}^i(\xi_2)=0, \quad
  n_-(\xi_2)=0,
  \label{eq:BD-3}
\end{equation}
where $i=1,2,\cdot\cdot\cdot,m$.
We have totally $2m+6$ boundary conditions 
(\ref{eq:BD-1})--(\ref{eq:BD-3})
for $2m+4$ unknown functions
$\Phi$, $E_\parallel$,
$n_\pm$, $g_\pm{}^1 \cdot\cdot\cdot g_\pm{}^m$. 
Thus two extra boundary conditions must be compensated 
by making the positions of the boundaries $\xi_1$ and $\xi_2$ be free.
The two free boundaries appear because $E_\parallel=0$ is imposed at the 
{\it both} boundaries and because $j_0$ is externally imposed.

\subsection{TeV Fluxes}

The pair production cascade is primarily described by the 
curvature-radiated, GeV $\gamma$-rays.
Therefore, the TeV emission due to Compton upscatterings 
is energetically negligible and
can be computed passively from the electrodynamic structure of the gap.
Assuming that the azimuthal gap width is 
$0.5 \pi W / \rlc (\equiv h)$ rad,
we can evaluate the ratio between the TeV flux and the infrared one as
(see \S 2.7 in Paper V)
\begin{eqnarray}
  f_{\rm R} 
  &\equiv&
  \frac{(\nu F_\nu)_{\rm TeV}}{(\nu F_\nu)_{0.1{\rm eV}}}
  < 5.63 \times 10^{-5} 
      \left( \frac{\rlc}{r_0} \sin\theta_0 \right)  
  \nonumber \\
  & & \hspace*{-1.0 truecm}
      \times \Gamma \left( j_0 \frac{D_\perp}{\rlc} \right)
             B_5
             \frac{\omega_{0.1{\rm eV}}}{\omega_{\rm TeV}} h^2,
  \label{eq:TeV_flux_2}
\end{eqnarray}
where $\omega_E$ refers to the solid angle in which photons with 
energy $E$ are emitted;
$\omega_{\rm TeV}=\omega_{0.1 \rm eV}$ is assumed for simplicity.
To estimate the upper bound, we take $D_\perp=\varpi_{\rm LC}$.
The inequality comes from the fact that the scattered photon energy
cannot exceed the particle energy.
It is noteworthy that it is the infrared photons with energy 
$\sim 0.1$ eV that contribute most effectively 
as the target photons of comptonization.
Neither the higher energy photons like surface blackbody X-rays 
nor the lower energy photons like polar-cap radio emission
contribute as the target photons,
because they have either too small cross sections 
or too small energy transfer when they are scattered. 

Substituting the X-ray fields, $\Omega$ and $\mu$ of individual pulsars
into the differential equations,
we can solve $\Ell(\xi)$ and hence $h$;
equation (\ref{eq:TeV_flux_2}) then gives
$f_{\rm R}$ as a function of $\inc$ and $j_0$.

\begin{figure} 
\centerline{ \epsfxsize=8.5cm \epsfbox[200 0 500 250]{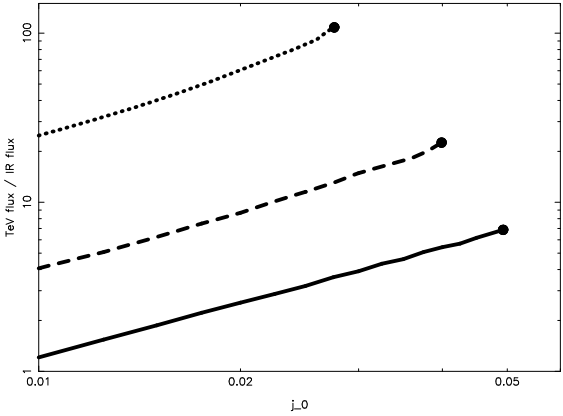} } 
\caption{\label{fig:fR} 
$f_{\rm R} \equiv (\nu F_\nu)_{\rm TeV} / (\nu F_\nu)_{0.1 \rm eV}$
vs. $j_0$ for B1509-58,
where $(\nu F_\nu)_E$ indicates the $\nu F_\nu$ flux at
photon energy $E$.
The solid, dashed, and doted lines correspond to $\inc=30^\circ$,
$45^\circ$, and $60^\circ$, respectively.
        }
\end{figure}

\section{Application to Individual Pulsars} 

In this section, we apply the theory to seven rotation-powered pulsars.
We first present the observed X-ray properties of individual pulsars
as input parameters in order of spin-down luminosity in table~1.
The X-ray spectrum of each pulsar is approximated
by separating into a soft blackbody component with temperature
$kT_{\rm s}$ and emitting area $A_{\rm s}$, 
a hard blackbody one with $kT_{\rm h}$ and $A_{\rm h}$, 
and a power-law one with
$dN_{\rm x}/d\Ex = N_{\rm pl} \Ex{}^\alpha$
($\epsilon_{\rm min}<\Ex<\epsilon_{\rm max}$),
where
$\epsilon_{\rm min}=0.1 \mbox{keV} / 511 \mbox{keV}$ and
$\epsilon_{\rm max}= 50 \mbox{keV} / 511 \mbox{keV}$
are assumed for homogeneous discussion.

\begin{table*}
  \centering
    \begin{minipage}{160mm}
      \caption{Input parameters and X-ray fields}
      \begin{tabular}{@{}lcccccccccl@{}}
        \hline
        \hline
        pulsar	
		& distance
		& $\Omega$		& $\log_{10}\mu$
		& $kT_{\rm s}$		& $A_{\rm s}/A_*{}^\dagger$
		& $kT_{\rm h}$		& $A_{\rm h}/A_*{}^\dagger$
		& $N_{\rm pl}$		& $-\alpha$
		& ref.	\\
        \	
		& kpc
		& rad s${}^{-1}$	& lg(G cm${}^3$)
		& keV			&
		& keV			& 
		& cm${}^{-3}$		&	
		&				\\
        \hline
        Crab	
		& 2.49
		& 188.1		& 30.53
		& $\ldots$	& $\ldots$
		& $\ldots$	& $\ldots$
		& $10^{17.30}$	& $1.8$	
		& 1				\\
        B0540-69
		& 49.4
		& 124.7		& 31.00
		& $\ldots$	& $\ldots$
		& $\ldots$	& $\ldots$
		& $10^{14.15}$	& $2.0$
		& 2				\\
        B1509-58
		& 4.40
		& 41.7		& 31.19
		& $\ldots$	& $\ldots$
		& $\ldots$	& $\ldots$
		& $10^{14.04}$	& $1.1$
		& 2				\\
        J1617-5055	
		& 3.30
		& 90.6		& 30.78
		& $\ldots$	& $\ldots$
		& $\ldots$	& $\ldots$
		& $10^{12.64}$	& $1.6$
		& 3,4				\\
        Vela	
		& 0.50
		& 61.3		& 30.53
		& 150		& 0.066
		& $\ldots$	& $\ldots$
		& $\ldots$	& $\ldots$
		& 5				\\
        B1951+32
		& 2.5
		& 159		& 29.68
		& $\ldots$	& $\ldots$
		& $\ldots$	& $\ldots$
		& $10^{13.55}$	& 1.6
		& 6				\\
        B1055-52
		& 1.53
		& 31.9		& 30.03
		& 68		& 7.3
		& 320		& $10^{-3.64}$
		& $\ldots$	& $\ldots$
		& 7				\\
        \hline
      \end{tabular}
      \begin{flushleft}
        ${}^\dagger$ 
	$A_*=4\pi r_*{}^2$ represents the whole neutron star surface area;
	$r_*=10$ km is assumed. \\
	References: 
        1~Knight (1982)					\qquad
	2~Saito (1998)					\qquad
        3~Torii et al. (1998)				\qquad
        4~Caswell et al. (1975)				\qquad
	5~$\ddot{\rm O}$gelman et al. (1993)		\qquad
        6~Safi-Harb and $\ddot{\rm O}$gelman (1995)	\qquad
	7~Greiveldinger et al. (1996)
      \end{flushleft}
    \end{minipage}
\end{table*}

Let us next consider the TeV fluxes.
As a typical example, we present
the results of $f_{\rm R}$ for B1509-58 as a function of 
$\inc$ and $j_0$ in figure~\ref{fig:fR}.
The filled circles correspond to the maximum current density $j_{\rm cr}$
above which no solution exists 
(see \S~3.1 in Paper III and \S~3.1 in Hirotani \& Okamoto 1998 
 for details)
if all the other parameters are fixed.
It follows from the figure that $f_{\rm R}$ becomes maximum at
$j_0=j_{\rm cr}$ (for a given $\inc$).
We thus consider the case of $j_0=j_{\rm cr}$
in this section to estimate the upper bound of $f_{\rm R}$.

The upper bound of $f_{\rm R}$ for the seven pulsars are
presented in table~2.
For Crab and B0540-69, the synchrotron (or synchro-curvature) process
dominates the simple curvature one when $\inc=60^\circ$;
therefore, this case is excluded in the table.
For B1055-52, acceleration length scale for a particle to attain
the saturated Lorentz factor
exceeds the gap half width for $\inc=60^\circ$.
In another word, the mono-energetic approximation breaks down;
therefore, the case of $\inc=60^\circ$ is excluded.
It follows from the table that
the TeV fluxes are unobservable ($< 10^{12}$ Jy Hz)
for moderate infrared fluxes ($< 10^{9}$ Jy Hz),
even when $j_0$ is adjusted close to $j_{\rm cr}$.

\noindent
\begin{table}
  \centering
    \begin{minipage}{140mm}
      \caption{Expected $\gamma$-ray properties}
      \begin{tabular}{@{}lcccc@{}}
        \hline
        \hline
        pulsar	
		& $\inc$
		& $j_0=j_{\rm cr}$
		& $W / \rlc{}^\dagger$
		& $f_{\rm R}{}^\ddagger$	\\
        \	& deg
		&
		&
		& 			\\
        \hline
        Crab	
		& 30
		& 0.0485
		& 0.083
		& $6.1 \times 10^{ 1}$	\\
	\ 	& 45
		& 0.0371
		& $0.064$
		& $1.8 \times 10^{ 2}$	\\
        B0540-69
		& 30
		& 0.0390
		& 0.074
		& $4.2 \times 10^{ 1}$	\\
	\ 	& 45
		& 0.0305
		& 0.053
		& $1.2 \times 10^{ 2}$	\\
        B1509-58
		& 30
		& 0.0375
		& 0.102
		& $5.1 \times 10^{ 0}$	\\
	\ 	& 45
		& 0.0399
		& 0.083
		& $2.3 \times 10^{ 1}$	\\
	\	& 60
		& 0.0276
		& 0.052
		& $1.1 \times 10^{ 2}$	\\
        J1617-5055
		& 30
		& 0.0627
		& 0.136
		& $4.5 \times 10^{ 1}$	\\
	\ 	& 45
		& 0.0430
		& 0.104
		& $1.2 \times 10^{ 2}$	\\
	\	& 60
		& 0.0241
		& 0.049
		& $3.2 \times 10^{ 2}$	\\
        Vela	
		& 30
		& 0.0620
		& 0.255
		& $2.8 \times 10^{ 1}$	\\
	\ 	& 45
		& 0.0375
		& 0.142
		& $4.2 \times 10^{ 1}$	\\
	\	& 60
		& 0.0264
		& 0.085
		& $1.7 \times 10^{ 2}$	\\
        B1951+32
		& 30
		& 0.0450
		& 0.133
		& $1.2 \times 10^{ 1}$	\\
	\ 	& 45
		& 0.0380
		& 0.095
		& $3.7 \times 10^{ 1}$	\\
	\	& 60
		& 0.0272
		& 0.057
		& $1.8 \times 10^{ 2}$	\\
        B1055-52
		& 30
		& 0.0343
		& 0.443
		& $2.1 \times 10^{ 0}$	\\
	\ 	& 45
		& 0.0232
		& 0.182
		& $1.6 \times 10^{ 0}$	\\
      \hline
      \end{tabular}
      \begin{flushleft}
        ${}^\dagger$ 
	$W$ indicates the full gap width. \\
        ${}^\ddagger$
	$f_{\rm R} \equiv (\nu F_\nu)_{\rm TeV} 
                          / (\nu F_\nu)_{0.1 \rm eV}$. \\
      \end{flushleft}
    \end{minipage}
\end{table}

\section{Discussion}

In summary, we have developed a one-dimensional model for 
an outer-gap accelerator.
As the inclination angle ($\inc$) increases,
the gap center (where $B_z$ vanishes) 
shifts starwards to be illuminated by a strong X-ray field,
which reduces the pair production mean
free path and hence the gap longitudinal width, $W$.
At the same time, the starwardly shifted gap has a strong magnetic field,
which increases the acceleration field (see eq. [\ref{eq:basic-2}]).
As a result, the TeV flux increases with increasing $\inc$,
even though $W$ decreases.
However, it is difficult to detect the TeV flux,
even when $\inc$ is as large as $60^\circ$.
We can therefore conclude that 
the difficulty of excessive TeV emission,
which appears in the picture of Cheng, Ho, and Ruderman (1986a,b),
does not arise in the present outer gap model.

It is seemingly possible to obtain observable TeV fluxes,
if $\inc$ exceeds $60^\circ$.
In this case, the \lq outer gap' is located close to the polar cap;
the distance from the star is within $13\%$ of $\rlc$.
The magnetic field there is so strong that
the synchrotron (or synchro-curvature) process becomes
important in general;
this could be verified by computing the pitch angle evolution of particles
created in the gap (see \S 5.3 in Paper I).
Such a \lq polar-gap-like' outer gap will be discussed in a separate paper.

For middle-age pulsars B0656+14 and Geminga, 
there is no stationary solution that satisfies the boundary conditions
presented in \S \ref{sec:BD}.
However, if we allow an external electric current 
flowing into the gap at the outer boundary, 
there exists a branch of solutions.
In this case, the gap is no longer located around the null surface
but shifts starwards.
Interestingly, when the external current
density approaches the Goldreich-Julian value,
the \lq outer gap' is found to shift to the polar cap 
and have a similar electrodynamic structure 
of Scharlemann, Arons, \& Fawley (1978).
Such a unification of the outer gap and polar cap accelerators
will be discussed in separate papers.

\acknowledgments

This research owes much to the helpful comments of Dr. S. Shibata.
The author wishes to express his gratitude to
Drs. Y. Saito and A. Harding for valuable advice. 
He also thanks the Astronomical Data Analysis Center of
National Astronomical Observatory, Japan for the use of workstations.


\begin{references}
\reference{bere89} 
  Berestetskii, V.\ B., Lifshitz, E.\ M.\ \& Pitaevskii, L.\ P., 1989, 
  Quantum Electrodynamics 3rd ed.
\reference{bes92} 
  Beskin, V.\ S., Istomin, Ya.\ N., \& Par'ev, V.\ I.\ 1992, 
  Sov. Astron. 36(6), 642
\reference{bowd87}
  Bowden, C. C. G. et al. 
  1993, Proc. of 23rd Int. Cosmic Ray Conf. (Calgary), 1, 294
\reference{casw75} 
  Caswell, J. L., Murray, J. D., Roger, R. S., Cole, D. J., 
  Cooke, D. J. 1975, A\& A 45, 239
\reference{cheA86} 
  Cheng, K. S., Ho, C.,   Ruderman, M., 1986a
  ApJ, 300, 500
\reference{cheB86} 
  Cheng, K. S., Ho, C.,   Ruderman, M., 1986b
  ApJ, 300, 522
\reference{edwa94} 
  Edwards, P. G., et al. 1994 
  A\& A 291, 468
\reference{grei96}
  Greiveldinger, C. et al. 
  1996, ApJ 465, L35
\reference{hiro99d}
  Hirotani, K. 1999 (Paper IV), submitted to ApJ
\reference{hiro98}
  Hirotani, K.   Okamoto, I.
  1998, ApJ 497, 563
\reference{hiro99a}
  Hirotani, K.   Shibata, S.,
  1999a (Paper I),  MNRAS 308, 54
\reference{hiro99b}
  Hirotani, K.   Shibata, S.,
  1999b (Paper II), MNRAS 308, 76
\reference{hiro99c}
  Hirotani, K.   Shibata, S.,
  1999c (Paper III), PASJ 51, 683
\reference{kifn96}
  Kifune, T. 1996, Space Science Reviews 75, 31
\reference{knig82}
  Knight F. K. 1982, ApJ 260, 538
\reference{nel93}
  Nel, H. I. et al. 
  1993, ApJ 418, 836
\reference{ogel93}
  $\ddot{\rm O}$gelman, H., Finley, J. P., Zimmermann, H. U. 
  1993, Nature 361, 136
\reference{roma96} 
  Romani, R. W. 1996,
  ApJ, 470, 469
\reference{safi95} 
  Safi-Harb, S. $\ddot{\rm O}$gelman, H.
  1995, ApJ 439, 722
\reference{safi95} 
  Safi-Harb, S. $\ddot{\rm O}$gelman, H.
  1995, ApJ 439, 722
\reference{sait98} 
  Saito, Y. 1998, Ph.D. Thesis
\reference{scha78} 
  Scharlemann, E. T., Arons, J., \& Fawley, W. M. 1978, ApJ, 222, 297
\reference{tori98}
  Torii, K. et al. 1998,
  ApJ 494, L207
\reference{usov94}
  Usov, V. V. 1994, ApJ 427, 394
\reference{yosh97}
  Yoshikoshi, T., et al.
  1997, ApJ 487, L65
\reference{zhan99}
  Zhang, L. Cheng, K. S. 1998, MNRAS 294, 177
\end{references}
\end{document}